\documentclass[
 reprint,
superscriptaddress,
showpacs,
amsmath,amssymb,
aps,
prl,
]{revtex4-1}

\usepackage{graphicx}
\usepackage{dcolumn}
\usepackage{bm}
\usepackage[colorlinks=true,linkcolor=blue,citecolor=blue,urlcolor=blue]{hyperref}
\usepackage{bm}
\usepackage{color}

\graphicspath{{Figures/Figure1/}{Figures/Figure2/}{Figures/Figure3/}{Figures/Figure4/}}

\begin{document}

\preprint{APS/123-QED}

\title{Cavity QED with magnetically coupled collective spin states}

\author{R. Ams\"{u}ss}
\affiliation{%
Vienna Center for Quantum Science and Technology, Atominstitut, TU Wien, 1020 Vienna, Austria\\
}%
\author{Ch. Koller}%
\affiliation{%
Vienna Center for Quantum Science and Technology, Atominstitut, TU Wien, 1020 Vienna, Austria\\
}%
\affiliation{Center for Micro- and Nanostructures ZNMS, TU Wien, 1040 Vienna, Austria}
\author{T. N\"{o}bauer}%
\author{S. Putz}%
\affiliation{%
Vienna Center for Quantum Science and Technology, Atominstitut, TU Wien, 1020 Vienna, Austria\\
}%
\author{S. Rotter}%
\affiliation{%
Institute for Theoretical Physics, TU Wien, 1040 Vienna, Austria\\
}
\author{K. Sandner}
\affiliation{%
Institute for Theoretical Physics, Universit\"{a}t Innsbruck, Technikerstrasse 25, 6020 Innsbruck, Austria\\
}
\author{S. Schneider}
\affiliation{%
Vienna Center for Quantum Science and Technology, Atominstitut, TU Wien, 1020 Vienna, Austria\\
}%
\author{M. Schramb\"{o}ck}
\affiliation{%
Vienna Center for Quantum Science and Technology, Atominstitut, TU Wien, 1020 Vienna, Austria\\
}%
\affiliation{Center for Micro- and Nanostructures ZNMS, TU Wien, 1040 Vienna, Austria}
\author{G. Steinhauser}
\affiliation{%
Vienna Center for Quantum Science and Technology, Atominstitut, TU Wien, 1020 Vienna, Austria\\
}
\author{H. Ritsch}
\affiliation{%
Institute for Theoretical Physics, Universit\"{a}t Innsbruck, Technikerstrasse 25, 6020 Innsbruck, Austria\\
}
\author{J. Schmiedmayer}%
\affiliation{%
Vienna Center for Quantum Science and Technology, Atominstitut, TU Wien, 1020 Vienna, Austria\\
}%
\author{J. Majer}%
\email{jmajer@ati.ac.at}
\affiliation{%
Vienna Center for Quantum Science and Technology, Atominstitut, TU Wien, 1020 Vienna, Austria\\
}%
\affiliation{Center for Micro- and Nanostructures ZNMS, TU Wien, 1040 Vienna, Austria}

\date{\today}

\begin{abstract}
We report strong coupling between an ensemble of nitrogen-vacancy center electron spins in diamond and a superconducting microwave coplanar waveguide resonator. The characteristic scaling of the collective coupling strength with the square root of the number of emitters is observed directly. Additionally, we measure hyperfine coupling to $^{13}\mathrm{C}$ nuclear spins, which is a first step towards a nuclear ensemble quantum memory. Using the dispersive shift of the cavity resonance frequency, we measure the relaxation time of the NV center at mK temperatures in a non-destructive way.
\end{abstract}


\maketitle

Building a practical quantum information processor will profit from a hybridization of different quantum systems in order to fulfill the requirements of long storage times, fast processing speeds, long-distance information transfer and scalability. A variety of hybrid systems have been conceived, including cold atoms and ions, nanomechanics, and molecules \cite{wallquist_hybrid_2009,verdu_strong_2009,andre_coherent_2006,rabl_hybrid_2006,twamley_superconducting_2010,marcos_coupling_2010}. Superconducting (SC) qubits \cite{clarke_superconducting_2008} are appealing processing devices due to the fast processing speed and their customizability. Superconducting coplanar waveguide (CPW) resonators have been used to reversibly transfer quantum states between two qubits and the resonator \cite{majer_coupling_2007}. Color centers in solids on the other hand -- most prominently the negatively charged nitrogen-vacancy defect in diamond (NV) -- show astonishingly long coherence times even at room temperature \cite{balasubramanian_ultralong_2009} and provide coherent bridges between electron spin resonances (ESR) in the GHz range and optical photons suitable for long-distance transfer \cite{togan_quantum_2010}.

The coupling between a single spin and the electromagnetic field is typically rather weak. However, when writing single excitations into ensembles of $N$ spins, it is enhanced by a factor $\sqrt{N}$ \cite{dicke_coherence_1954}. Provided that excitations can be coherently written to and retrieved from the ensemble by strong coupling to high-Q resonators, quantum memories can be constructed \cite{imamoglu_cavity_2009,wesenberg_quantum_2009}. Efforts parallel to this work have succeeded in demonstrating strong ensemble-resonator coupling using NVs \cite{kubo_strong_2010} and other impurity spins \cite{schuster_high-cooperativity_2010,bushev_rare_2011,chiorescu_magnetic_2010}. Schemes such as gradient echoes \cite{alexander_photon_2006} and controlled reversible inhomogeneous broadening \cite{kraus_quantum_2006} can be used to store quantum information \cite{wu_storage_2010}.
\begin{figure}[b]
\includegraphics{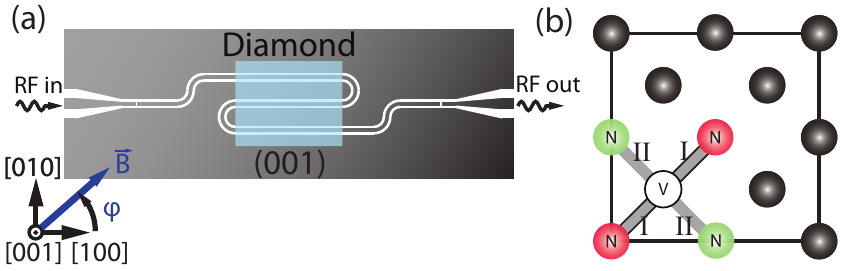}
\caption{\label{fig1} (a) (Color online) Schematic of the experimental setup. A (001) diamond is positioned in the middle and parallel  to the surface of a $\lambda/2$ CPW resonator. The chip dimensions are $12 \times 4$~mm, the center conductor is 20~$\mathrm{\mu m}$ wide, the gaps are 8.3~$\mathrm{\mu m}$. A 2-axis Helmholtz coil configuration creates a magnetic field in an arbitrary direction $\varphi$ within the (001) plane, tuning the NV centers. (b) Sketch of a diamond lattice with the four possible NV center orientations. Due to the diamond symmetry, the NV ensembles denoted as I (red) and II (green) are tuned equivalently, since these ensembles enclose the same angle with the applied magnetic field.}
\end{figure}
\begin{figure*}[!t]
\includegraphics[scale=0.95]{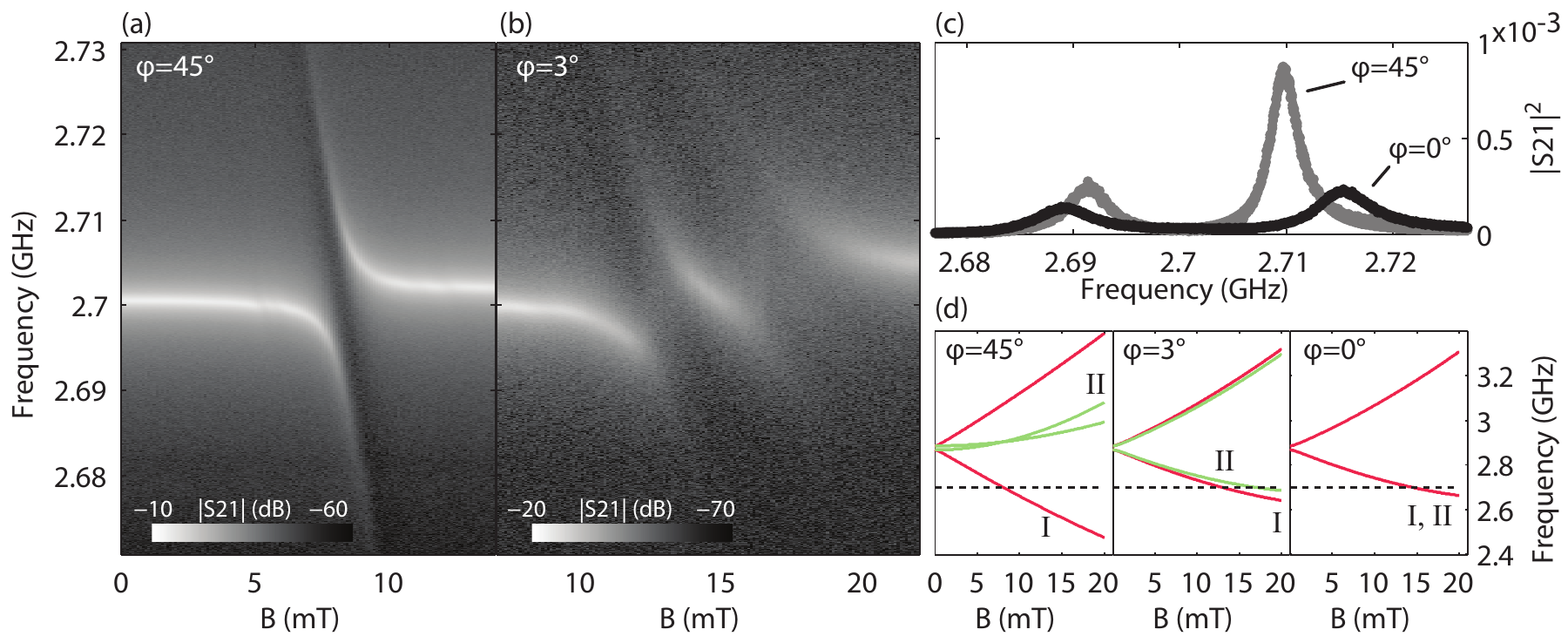}
\caption{\label{fig2}(Color online) Resonator transmission $|S_{21}|^2$ as a function of the magnetic field for (a) $\varphi=45^{\circ}$ and (b) $\varphi=3^{\circ}$. While in (a) an avoided crossing of the coupled resonator-spin ensemble system is observed when $\omega_r=\omega_-^I$, we see in (b) the additional splitting of the $\omega_- ^{II}$ transition. (c) Comparison of the vacuum Rabi splitting of $2g_I/2\pi=18.5$~MHz for $\varphi=45^{\circ}$ ($B=8.2$~mT) (gray line) and the $\sqrt{2}$ larger splitting of $2g/2\pi=26.3$~MHz (black line) for $\varphi=0^{\circ}$ ($B=14.3$~mT). Each dataset is fitted by a two-peak Lorentzian curve. (d) Transition frequencies $\omega_{\pm}^{I}$ and $\omega_{\pm}^{II}$ between the ground $\vert0\rangle_{I,II}$ and the excited states $\vert \pm\rangle_{I,II}$ for different magnetic field angles ($\varphi=45^{\circ},3^{\circ},0^{\circ}$).The dashed lines indicate the cavity frequency.}

\end{figure*}

In this Letter, we employ a strongly coupled system consisting of a CPW resonator and an ensemble of NVs to directly demonstrate key features of ensemble coupling, such as the $\sqrt{N}$ scaling of the coupling rate. Properties of the spin ensemble including the NV spin-lattice relaxation time $T_1$  and  ESR zero-field splitting parameters, are determined via the cavity resonance shift caused by dispersive interactions. We also observe the hyperfine interaction between NVs and nearby $^{13}\mathrm{C}$ nuclear spins.

The experimental setup is sketched in Fig.~\ref{fig1}: A $4.5 \times 2.25 \times 0.5$~mm$^3$ high-temperature high-pressure diamond is placed on top of a superconducting coplanar waveguide resonator and fixated via a spring-loaded metallic pin. Our single crystal diamond sample has two polished (100) surfaces and is specified to contain a nitrogen impurity concentration of about 200~ppm. In order to form negatively charged NV defect centers, the sample was neutron irradiated with a fluence of $5\times10^{17}\:\mathrm{cm^{-2}}$ to create vacancies \cite{mita_change_1996}. Subsequent annealing for four hours at $900\,^{\circ}\mathrm{C}$ resulted in an NV concentration of about 6~ppm which was determined by comparing the fluorescence intensity of an ensemble to a single NV.
Resonator chip and diamond are enclosed in a copper box and positioned in the center of two pairs of perpendicular Helmholtz coils. This coil configuration allows to apply a homogeneous magnetic field~$B$ (up to about 25~mT) over the extent of the diamond and allows for adjusting the magnetic field direction in the plane of the resonator surface, without changing the resonators frequency \cite{healey_magnetic_2008}.

To explain our experimental observations, we consider the spin Hamiltonian of a single NV center \cite{jelezko_observation_2004}
\begin{eqnarray}
H_{NV} = g_e \beta_e \mathbf{B} \cdot \mathbf{S} + \mathbf{S} \overset{\text{\tiny$\leftrightarrow$}}{\textbf{D}} \mathbf{S} - g_n \beta_n \mathbf{B} \cdot \mathbf{I} + \mathbf{I} \overset{\text{\tiny$\leftrightarrow$}}{\mathbf{A}} \mathbf{S},
\label{eqn1}
\end{eqnarray}
where the first and third term describe the Zeeman interaction of an electronic NV spin ($S=1$, $g_e=2.0028$) and a $^{13}\mathrm{C}$ nuclear spin  ($I=1/2$, $g_n=1.4048$) with an external magnetic field $\mathbf{B}$, the second the zero-field splitting (ZFS) ($D/h=2.877$~GHz, $E/h=7.7$~MHz) of the NV ground state and the last the hyperfine interaction of the electron and nuclear spin (hyperfine coupling parameters $A_{\parallel}/h=200$~MHz, $A_{\perp}/h=120$~MHz). The Bohr magneton and the nuclear magneton are denoted as $\beta_e$ and $\beta_n$. The eigenenergy spectrum was obtained by diagonalizing the Hamiltonian and confirmed with \textsc{EasySpin} \cite{stoll_easyspin_2006}.

Although the main axis of an NV center is along any of the four crystallographic $\langle$111$\rangle$ directions, for the (001) crystal plane there are only two magnetically inequivalent directions, labeled I and II in Fig.~\ref{fig1}(b). Hence, applying a magnetic field in an arbitrary direction within the (001) plane divides the whole ensemble of NV centers into two magnetic subensembles due to the electron Zeeman effect. Consequently, each subensemble will be tuned into resonance with the CPW resonator at a different magnetic field value (see Fig.~\ref{fig2}(d)). 
For the CPW resonator, we chose a capacitively coupled $\lambda/2$ niobium SC-resonator, fabricated on sapphire substrate. A meander geometry ensures a large overlap of the magnetic mode volume with the spin ensemble. At zero magnetic field and with the diamond on top, the resonator has a center frequency of $\omega_r/2\pi= 2.701$~GHz and a quality factor $Q \approx 3200$. Comparing these values with the empty and coupling limited resonator ($\omega_r/2\pi= 2.897$~GHz and  $Q \approx 4\times 10^4$) shows that the diamond introduces considerable losses, possibly due to surface conductivity effects \cite{maier_origin_2000}. All subsequently shown measurements of the resonator $S_{21}$-parameter are performed by using a vector network analyzer with a probe power corresponding to roughly $8\times10^4$ photons in the resonator. Lowering the power did not alter the signal, except for lowering the signal to noise ratio \cite{gripp_anharmonicity_1996}. During all measurements the temperature was below 60~mK, ensuring a high degree of spin-polarization and a low thermal photon background in the frequency range of interest.

A calculation similar to Verdu  \emph {et al.} \cite{verdu_strong_2009} adapted for NV centers (see also Kubo \emph{et al.} \cite{kubo_strong_2010}) shows that the small mode volume of a CPW resonator leads to a coupling strength of $g_0/2\pi\approx$ 12 Hz for a single NV center 1~$\mathrm{\mu}$m above the center conductor. This is not enough to exceed the losses  $\kappa/2\pi\approx0.8$~MHz  of the resonator and the decoherence rate of the color center. However, by coupling a large number of spins $N\approx10^{12}$ collectively to the same electromagnetic mode of the resonator, an enhancement of $g_{col}/2\pi\approx\sqrt{N} g_{0}/2\pi \approx 10$~MHz allows to enter the strong coupling regime. This can be modeled by a generalized Hamiltonian
\[
H = \hbar\omega_r a^{\dagger} a+ \frac{\hbar}{2} \sum_{i,j} \omega_{ij}\sigma^z_{ij} +  \hbar \sum_{i,j}g_{ij} (\sigma^+_{ij}a+a^{\dagger} \sigma^-_{ij}),
\label{eqn2}
\]
where the index $i$ runs over the different Zeeman subensembles I, II and $j$ accounts for an inhomogeneous distribution of spin transition frequencies $\omega_{ij}$ as well as coupling strengths $g_{ij}$. The operators $\sigma^+_{ij}$ and $\sigma^-_{ij}$ describe the spin raising and lowering operators, $a^{\dagger}$ and $a$ denote the creation and annihilation of a cavity photon. Although the presence of inhomogeneous broadening of the spin transitions has interesting consequences \cite{diniz_strongly_2011,kurucz_spectroscopic_2011}, the subsequent analysis assumes $\omega_{ij}=\omega_{i}$.  The effective coupling strength then reduces to $g_{col}=\sqrt{\sum_i g_i^2}$. Furthermore, if the number of excitations meets $n_{exc}\ll\sqrt{N}$, the spin ensemble is well approximated by a system of coupled harmonic oscillators \cite{henschel_cavity_2010}.

For strong coupling, an avoided crossing with a level-splitting of $2g_{col}$ is expected when the ensemble of spins is tuned into resonance with the cavity. Fig.~\ref{fig2}(a) shows the transmission through the resonator as the magnetic field amplitude is varied for a field direction of $\varphi=45^{\circ}$. There, a large Rabi splitting of $2g_{I}/2\pi=18.5$~MHz is observed, corresponding to a single Zeeman subensemble in resonance ($\omega_r=\omega_{-}^{I}$). The FWHM linewidths of the two peaks are found to be slightly different $\approx 2 \pi \times (4.4,3.3)$~MHz at the avoided crossing.
By applying a field in the [100] direction ($\varphi=0^{\circ}$), where the energy levels of the two subensembles I and II coincide, we can directly measure the $\sqrt{N}$-enhancement. Since twice as many spins couple to the field mode, an enhancement compared to the case for $\varphi=45^{\circ}$ of $g_{col}\approx\sqrt{2}g_{I}$  is expected. Indeed we obtain an enhancement by a factor of 1.42 ( $2 g_{col}/2\pi=26.3$~MHz). We also note an increase of the linewidth of the two peaks to $\approx 2\pi\times7.3$~MHz for $\varphi=0^{\circ}$. In case the magnetic field projections experienced by the two NV subensembles do not coincide (e.g. $\varphi=+3^{\circ}$), two distinct avoided crossings are observed (see Fig.~\ref{fig2}(b), (d)).

Qualitatively one can understand the observed difference in peak amplitude and peak width in Fig.~\ref{fig2}(c) by the presence of higher lying NV states ( $\vert + \rangle_{I} $, $\vert - \rangle_{II} $). Including these states leads to a model of three (or more) coupled oscillators, where the peaks do not have equal height and width \cite{garraway_nonperturbative_1997}. Furthermore inhomogeneous broadening, which is not contained in a model of coupled oscillators, influences the width and shape of the peaks \cite{kurucz_spectroscopic_2011}.

\begin{figure}
\includegraphics{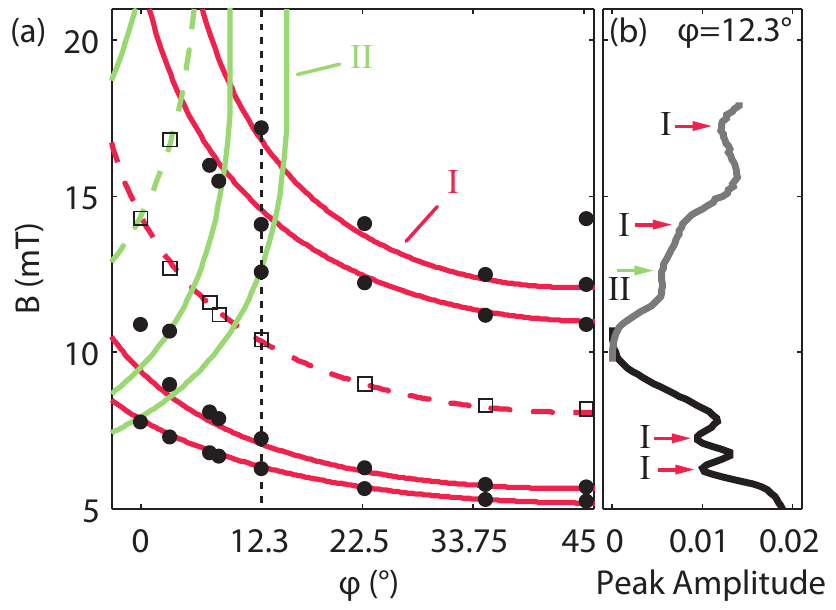}
\caption{\label{fig4}(Color online) (a) Measured (dots) and simulated (solid lines) nearest neighbor $^{13}\mathrm{C}$ satellite lines as a function of the magnetic field angle (colors denote the subensembles I and II). Also shown is the measured (center of the avoided crossing) and theoretical position of the main NV ESR (squares and dashed lines). The dotted line marks the angle at which the data in (b) is taken. (b) Amplitude of the lower (black) and higher (gray) frequency peak of the transmission signal obtained from Lorentzian fits as a function of the magnetic field. Satellite lines due to the hyperfine interaction of nearest neighbor $^{13}\mathrm{C}$ nuclear spins with the electronic NV are visible and indicated by arrows. For $\varphi=12.3^{\circ}$ we see the four allowed ESR transitions of subensemble I (red arrow) and one of subensemble II (green arrow).}
\end{figure}

Another interesting observation is the appearance of weak satellite lines around the main ESR line, visualized in Fig.~\ref{fig4}(b) by the resonator transmission amplitude as a function of the magnetic field. As previously shown \cite{jelezko_observation_2004}, the hyperfine interaction of a nearest neighbor $^{13}\mathrm{C}$ nuclear spin with the NV spin leads to an ESR doublet split by 130~MHz in zero field. However, for our geometry and a non-zero magnetic field, the pseudo-nuclear Zeeman effect \cite{bleaney__1967} gives rise to two additional transitions, resulting in four allowed ESR lines (selection rules $|\Delta m_S| =1$, $|\Delta m_I|=0$). In Fig.~\ref{fig4}(a) we show the position of the satellite lines as a function of the magnetic field direction. These findings agree well with the simulation of the full spin Hamiltonian Eq.~\ref{eqn1} when we assume hyperfine coupling parameters $A_{\parallel}/h=200$~MHz and $A_{\perp}/h=120$~MHz \cite{felton_hyperfine_2009}. Since the isotope $^{13}\mathrm{C}$ only appears at its natural abundance of 1.1$\%$ in our samples, we estimate the coupling strength of these resonances to be weaker by a factor of $\sqrt{3\times1.1/100}$. For this reason we do not observe an avoided crossing for these lines but only a reduction of the transmission amplitude and an increased linewidth. Also note that some remote satellite lines observed are not predicted by the model.

\begin{figure}
\includegraphics{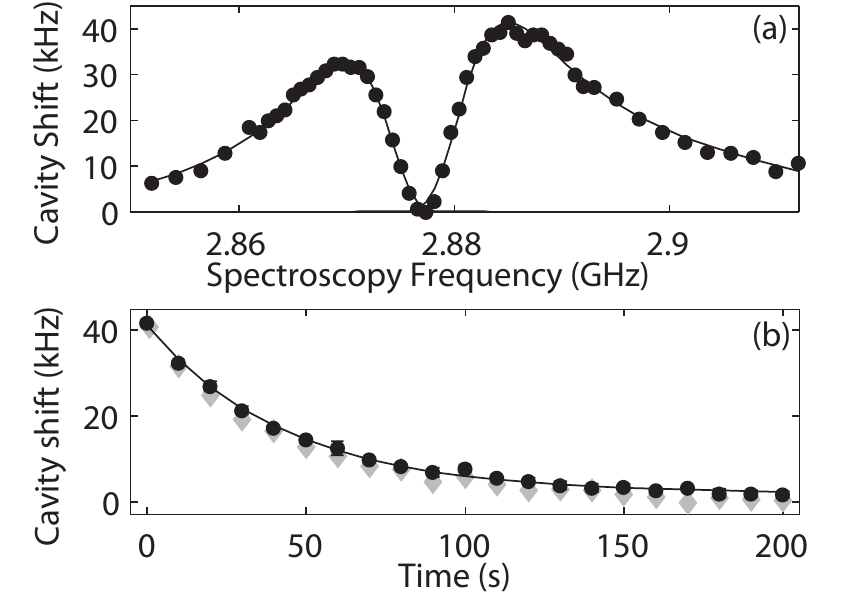}
\caption{\label{fig3} (a) Dispersive resonator shift as a function of the spectroscopy frequency $\omega_d/2\pi$ ($B=0$~mT). This measurement allows to extract the zero-field splitting parameters $D/h=2.877$~GHz and $E/h=7.7$~MHz. (b) NV spin relaxation measured via the resonator shift after the excitation frequency has been switched off  ($B=0$~mT, $\omega_d/2\pi=2.885$~GHz). Black dots refer to resonator frequency measurements after a given time after switch off ($\tau=44\pm4$~s), diamonds mark measurements with a continuously probed resonator ($\tau=45\pm4$~s) (see text for details). An exponential fit of the black data points is also displayed. }
\end{figure}

Additionally, we use a cavity readout scheme in the dispersive regime of cavity QED, where $\Delta=|\omega_r-\omega_{\pm}|\gg \sqrt{N}g$. Here the atomic ensemble can be seen as a refractive medium, lowering the resonator frequency by $\Delta_c\simeq Ng^2/\Delta$ when the ensemble is polarized in the ground state. If we add an additional MW tone $\omega_d=\omega_{\pm}$, a fraction of the NV ensemble is excited, leading to a positive shift which is plotted in figure \ref{fig3}. In Fig.~\ref{fig3}(a) we scan the spectroscopy frequency $\omega_d$ across the NV ESR lines and obtain the zero-field splitting parameters of the diamond sample by making use of the aforementioned dispersive interaction. We extract values for the parallel component of the zero-field splitting $D/h=2.877$~GHz and the transversal component $E/h=7.7$~MHz from the position of the minimum and the distance of the two peak maxima, respectively. As a fit function we used a simple asymmetric lineshape \cite{stancik_simple_2008}. These values are in accordance with ODMR measurements of the same sample and taking temperature effects into account \cite{acosta_diamonds_2009}. 

We also present a relaxation measurement of the spin ensemble after excitation at $\omega_d/2\pi =2.885$~GHz (see Fig.~\ref{fig3}(b)). In this measurement we first excite the NV centers, then wait for a given time after the dispersive tone has been switched off and finally probe the cavity resonance. This sequence is repeated for different waiting times. The resonant cavity probe signal is switched on only in this last part of the measurement sequence. From an exponential fit we determine a relaxation time constant $\tau=44\pm4$~s. This value agrees well with previous T1 measurements of NV centers in diamond at low temperature \cite{harrison_measurement_2006,jarmola__????}. In order to study the influence of the probe signal on the spin relaxation, we excite the NV centers as before and then probe the cavity continuously on resonance. For low probe powers ( $\approx 5 \times 10^4$ photons) the obtained time constant $\tau=45\pm4$~s agrees with the previously measured decay time. This shows that the measurement is non-destructive which is due to the fact that the dispersive measurement is a quantum non-demolition measurement at low probe power. However, we note that for higher probe powers the observed time constant substantially decreases and the form of the decay deviates from a simple monoexponential decay.

In conclusion we have shown the strong coupling of an NV spin ensemble to a CPW resonator and its $\sqrt{N}$-scaling with the number of emitters. We determined the spin-lattice relaxation time and ZFS parameters at 60~mK using dispersive measurements, a powerful technique in cavity QED. Furthermore the observed coupling to nuclear spin degrees of freedom in diamond may provide an avenue towards extended storage times and quantum information processing with auxiliary nuclear spins. 

\begin{acknowledgments}
This work was supported by the the Austrian FWF through the Wittgenstein Prize and the EU (MI-DAS, AQUTE). R.A., T.N. were supported by CoQuS,  C.K. by FunMat and, K.S. the DOC-fFORTE doctoral program. We thank J. Wrachtrup, F. Jelezko, H. Fedder and M. Trupke for helpful discussions.
\end{acknowledgments}
%

\end{document}